\documentclass[aps,pra,showpacs]{revtex4}
\usepackage{graphicx}
\usepackage{amssymb}

\newcommand*{\Tr}{\mathop{\mathrm{Tr}}} 
\newcommand*{\re}{\mathop{\mathrm{Re}}}

\begin{document}

\title{Quantum Trajectory method for the Quantum Zeno and anti-Zeno effects}

\author{J. Ruseckas}

\email{ruseckas@itpa.lt}

\homepage{http://www.itpa.lt/~ruseckas}

\affiliation{Institute of Theoretical Physics and Astronomy of Vilnius
University\\ A. Go\v{s}tauto 12, LT-01108 Vilnius, Lithuania}

\author{B. Kaulakys}

\affiliation{Institute of Theoretical Physics and Astronomy of Vilnius
University\\ A. Go\v{s}tauto 12, LT-01108 Vilnius, Lithuania}

\begin{abstract}
We perform stochastic simulations of the quantum Zeno and anti-Zeno effects for
two level system and for the decaying one. Instead of simple projection
postulate approach, a more realistic model of a detector interacting with the
environment is used. The influence of the environment is taken into account
using the quantum trajectory method. The simulation of the measurement for a
single system exhibits the probabilistic behavior showing the collapse of the
wave-packet. When a large ensemble is analysed using the quantum trajectory
method, the results are the same as those produced using the density matrix
method. The results of numerical calculations are compared with the analytical
expressions for the decay rate of the measured system and a good agreement is
found. Since the analytical expressions depend on the duration of the
measurement only, the agreement with the numerical calculations shows that other
parameters of the model are not important.
\end{abstract}


\maketitle

\section{Introduction}

The quantum Zeno effect has attracted much attention. The effect is caused by
the influence of the measurement on the evolution of a quantum system. This
effect is related to the non-exponential survival probability of the quantum
system.

The exponential survival law is known to be an excellent phenomenological fit
to unstable phenomena. However, from quantum mechanics it follows that the
survival probability is not exponential for short and long times. Short-time
behavior of the survival probability is not exponential but quadratic
\cite{khalfin58}. The deviation from the exponential decay has been confirmed
by Wilkinson \emph{et al.} \cite{wilkinson97}. This result, when combined with
the frequent measurements, leads to what is known as the quantum Zeno effect
\cite{mishra77}. Nowadays there are a number of experiments which claim to have
verified the quantum Zeno effect and some others are planned
\cite{itano90,kwiat95,nagels97,toschek01}. It was also predicted that frequent
measurements could accelerate the decay process
\cite{kaulakys97,kofman00,lewenstein00,facchi01PRL,ruseckas01,kofman01}. This
is the so-called quantum anti-Zeno effect. Both effects were first observed in
an atomic tunneling experiment \cite{fisher01}.

The states of the system need not to be frozen: in the general situation the
coherent evolution of the system can take place in dynamically generated
quantum Zeno subspaces \cite{facchi02}. The projective measurements used in the
description of the quantum Zeno effect can be replaced by another quantum
system interacting strongly with the principal one
\cite{facchi01,simonius78,harris82,schulman98}.

Interaction with the measuring device is one of many possible interactions of
the system with an external environment.  It is known that not only measurements
cause consequences similar to the quantum Zeno effect on the system's evolution
\cite{panov99,mensky99}. The experiment of Itano \emph{et al.} \cite{itano90}
has been explained in Refs.~\cite{petrosky90,frerichs91,pascazio94} using the
dynamical description, without using the concept of the measurements.  It was
shown that the quantum Zeno effect follows from the quantum theory of
irreversible processes, as well. Therefore, the quantum Zeno and anti-Zeno
effects can be considered as more general phenomena. However, in this paper we
will consider only one particular phenomenon, i.e., the effects caused by the
measurements.

Quantum Zeno effect can have practical significance in quantum computing. The
use of the quantum Zeno effect for correcting the errors in quantum computers
was first suggested by Zurek \cite{zurek84}. A number of quantum codes
utilising the error prevention that occurs in the Zeno limit have been proposed
\cite{vaidman96,beige04,erez04,franson04,brion05}.

In the analysis of the quantum Zeno effect the projection postulate is not
sufficient. The measurement should be described more fully, including the
detector into the description. In the description of the quantum measurement
process, the environmentally induced decoherence plays a very important role
\cite{zurek81,zurek82,walls85,unruh89,zurek91,walls94,giuilini96}. Therefore,
in order to correctly describe the measurement process one should include into
the description the interaction of the detector with the environment. In this
paper we describe the evolution of the detector interacting with the
environment using the quantum jump model developed by Carmichael
\cite{carmichael93}.

The density matrix analysis assumes that the experiment is performed on a large
number of systems. An alternative to the density matrix analysis are stochastic
simulation methods
\cite{dalibard92,gisin92,carmichael93,hegerfeldt93,moelmer93}. Various
stochastic simulation methods describe quantum trajectories for the states of
the system subjected to random quantum jumps. Using stochastic methods one can
examine the behaviour of individual trajectories, therefore such methods
provide the description of the experiment on a single system in a more direct
way. The results for the ensemble are obtained by repeating the stochastic
simulations several times and calculating the average.

Stochastic simulations of the quantum Zeno effect experiment were performed in
Ref.~\cite{power96}. In present paper we use the quantum jump method to
describe the evolution of frequently measured systems and to compare the
numerical results with the analytically obtained decay rates.

The paper is organized as follows. In Sec.~\ref{sec:description} we present the
description of the measurement. The model of the detector is presented in
Sec.~\ref{sec:detector}. In Sec.~\ref{sec:stochastic-methods} we present the
quantum jump method briefly. In Sec.~\ref{sec:stochastic-simulation} evolution
of the detector is calculated using the quantum jump method. Evolution of
frequently the measured two level system is investigated in
Sec.~\ref{sec:measured-two-level}. In Sec.~\ref{sec:decaying-system} we present
a numerical model of the decaying system. Using this model in
Sec.~\ref{sec:measurement-decaying} we calculate the evolution of the
frequently measured decaying system. Section \ref{sec:concl} summarizes our
findings.

\section{\label{sec:description}Description of the measured system}

We consider a system that consists of two parts: $A$ and $F$. The system $A$ is
interacting with the detector, i.e., it is measured. We assume that the system
$A$ has the discrete energy spectrum. The Hamiltonian of this part is
$\hat{H}_A$. The other part of the system is represented by Hamiltonian
$\hat{H}_F$. Hamiltonian $\hat{H}_F$ commutes with $\hat{H}_A$. The operator
$\hat{V}$ causes the jumps between different energy levels of $\hat{H}_A$.
Therefore, the full Hamiltonian of the system is of the form
$\hat{H}_S=\hat{H}_A+\hat{H}_F+\hat{V}$. The example of such a system is an
atom with the Hamiltonian $\hat{H}_A$ interacting with the electromagnetic
field, represented by $\hat{H}_F$, while the interaction between the atom and
the field is $\hat{V}$.

In this article we consider the system $A$ that has two levels: ground
$|g\rangle$ and excited $|e\rangle$. We will measure whether the system is in
the ground state. The measurement is performed by coupling the system with the
detector. The full Hamiltonian of the system and the detector equals to 
\begin{equation}
\hat{H}=\hat{H}_S+\hat{H}_D+\hat{H}_I,
\label{eq:ham}
\end{equation}
where $\hat{H}_D$ is the Hamiltonian of the detector and $\hat{H}_I$ represents
the interaction between the detector and the measured system, described by the
Hamiltonian $\hat{H}_A$. We can choose the basis
$|n\alpha\rangle=|n\rangle\otimes|\alpha\rangle$ common for the operators
$\hat{H}_A$ and $\hat{H}_F$, 
\begin{eqnarray}
\hat{H}_A|n\rangle & = &\hbar\omega_n|n\rangle,
\label{eq:HA}
\\
\hat{H}_F|\alpha\rangle & = &\hbar\omega_{\alpha}|\alpha\rangle.
\end{eqnarray}
Here $\hbar\omega_n$ and $\hbar\omega_{\alpha}$ are energies of the systems $A$
and $F$, respectively.

The initial density matrix of the system is $\hat{\rho}_S(0)$. The initial
density matrix of the detector is $\hat{\rho}_D(0)$. Before the measurement the
measured system and the detector are uncorrelated, therefore, the full density
matrix of the measured system and the detector is
$\hat{\rho}(0)=\hat{\rho}_S(0)\otimes\hat{\rho}_D(0)$.

When the interaction of the detector with the environment is taken into
account, the evolution of the measured system and the detector cannot be
described by a unitary operator. More general description of the evolution,
allowing to include the interaction with the environment, can be given using
the superoperators. Therefore, we will assume that the evolution of the
measured system and the detector is given by the superoperator $\mathcal{S}(t)$.

\subsection{\label{sec:meas-unpert}Measurement of the unperturbed system}

In this section we investigate the measurement of the unperturbed system, i.e.,
the case when $\hat{V}=0.$

We assume that the Markovian approximation is valid i.e., the evolution of the
measured system and the detector depends only on their state at the present
time. The master equation for the full density matrix of the detector and the
measured system is
\begin{equation}
\frac{\partial}{\partial
t}\hat{\rho}(t)=\frac{1}{i\hbar}[\hat{H}_A,\hat{\rho}(t)]+\frac{1}{i\hbar}[\hat{
H}_I,\hat{\rho}(t)]+\frac{1}{i\hbar}[\hat{H}_D,\hat{\rho}(t)]+\mathcal{L}_D\hat{
\rho}(t),
\label{eq:master-unp}
\end{equation}
where the superoperator $\mathcal{L}_D$ accounts for the interaction of the
detector with the environment. We assume that the measurement of the
unperturbed system is a quantum non-demolition measurement
\cite{unruh79,caves80,braginsky80,braginsky96}. The measurement of the
unperturbed system does not change the state of the measured system when
initially the system is in an eigenstate of the Hamiltonian $\hat{H}_A$. This
can be if $[\hat{H}_A,\hat{H}_I]=0$.

We introduce the superoperator $\mathcal{L}_{n,m}$ acting only on the density
matrix of the detector
\begin{equation}
\mathcal{L}_{n,m}\hat{\rho}=\frac{1}{i\hbar}(\langle
n|\hat{H}_I|n\rangle\hat{\rho}-\hat{\rho}\langle
m|\hat{H}_I|m\rangle)+\frac{1}{i\hbar}[\hat{H}_D,\hat{\rho}]+\mathcal{L}_D\hat{
\rho}
\label{eq:lnm}
\end{equation}
and the superoperator $\mathcal{S}_{n,m}(t)$ obeying the equation
\begin{equation}
\frac{\partial}{\partial
t}\mathcal{S}_{n,m}(t)=\mathcal{L}_{n,m}\mathcal{S}_{n,m}(t),
\end{equation}
with the initial condition $\mathcal{S}_{n,m}(0)=1$. Then the full density
matrix of the detector and the measured system after the measurement is
\begin{equation}
\hat{\rho}(\tau_M)=\mathcal{S}(\tau_M)\hat{\rho}(0)=\sum_{
n,m}|n\rangle(\rho_S)_{n,m}(0)e^{i\omega_{m,n}\tau_M}\langle
m|\otimes\mathcal{S}_{n,m}(\tau_M)\hat{\rho}_D(0),
\label{eq:full}
\end{equation}
where $\tau_M$ is the duration of the measurement and
\begin{equation}
\omega_{m,n}=\omega_m-\omega_n
\end{equation}
with $\omega_n$ defined by Eq.~(\ref{eq:HA}). From Eq.~(\ref{eq:full}) it
follows that the non-diagonal matrix elements of the density matrix of the
system after the measurement
$(\rho_S)_{n,m}(\tau_M)\equiv(\rho_S)_{n,m}(0)e^{i\omega_{m,n}\tau_M}$ are
multiplied by the quantity
\begin{equation}
F_{n,m}(\tau_M)\equiv\Tr\{\mathcal{S}_{n,m}(\tau_M)\hat{\rho}_D(0)\}.
\label{eq:F}
\end{equation}
Since after the measurement the non-diagonal matrix elements of the density
matrix of the measured system should become small (they must vanish in the case
of an ideal measurement), $F_{n,m}(\tau_M)$ must be also small when $n\neq m$.

\section{\label{sec:detector}The detector}

We take an atom with two energy levels, the excited level $|a\rangle$ and the
ground level $|b\rangle$, as the detector. The Hamiltonian of the detecting
atom is
\begin{equation}
\hat{H}_D=\frac{\hbar\Omega_D}{2}\hat{\sigma}_z.
\end{equation}
Here $\hbar\Omega_D$ defines the separation between levels $|a\rangle$ and
$|b\rangle$, $\hat{\sigma}_x$, $\hat{\sigma}_y$, $\hat{\sigma}_z$ are Pauli
matrices. The interaction Hamiltonian $\hat{H}_I$ we take as
\begin{equation}
\hat{H}_I=\hbar\lambda|g\rangle\langle g|(\hat{\sigma}_{+}+\hat{\sigma}_{-}),
\end{equation}
where $\hat{\sigma}_{\pm}=\frac{1}{2}(\hat{\sigma}_x\pm i\hat{\sigma}_y)$. The
parameter $\lambda$ describes the strength of the coupling with the detector.
The detecting atom interacts with the electromagnetic field. The interaction of
the atom with the field is described by the term
\begin{equation}
\mathcal{L}_D\hat{\rho}_D=-\frac{\Gamma}{2}(\hat{\sigma}_{+}\hat{\sigma}_{
-}\hat{\rho}_D-2\hat{\sigma}_{-}\hat{\rho}_D\hat{\sigma}_{+}+\hat{\rho}_D\hat{
\sigma}_{+}\hat{\sigma}_{-}),
\label{eq:LD}
\end{equation}
where $\Gamma$ is the atomic decay rate.

At the equilibrium, when there is no interaction with the measured system, the
density matrix of the detector is $\hat{\rho}_D(0)=|b\rangle\langle b|$.

\subsection{Duration of measurement}

We can estimate the characteristic duration of one measurement $\tau_M$ as the
time during which the non-diagonal matrix elements of the measured system
become negligible. Therefore, in order to estimate the duration of the
measurement $\tau_M$, we need to calculate the quantity
\[
F_{e,g}(t)=\Tr\{\mathcal{S}_{e,g}(t)\hat{\rho}_D(0)\}.
\]
We will solve the equation
\begin{equation}
\frac{\partial}{\partial
t}\hat{\rho}_D=\frac{1}{i\hbar}[\hat{H}_D,\hat{\rho}_D]+\mathcal{L}_{e,g}\hat{
\rho}_D.
\end{equation}
For the matrix elements of the density matrix of the detector we have the
equations
\begin{eqnarray}
\frac{\partial}{\partial t}\rho_{aa} & = & i\lambda\rho_{ab}-\Gamma\rho_{aa},\\
\frac{\partial}{\partial t}\rho_{bb} & = & i\lambda\rho_{ba}+\Gamma\rho_{aa},\\
\frac{\partial}{\partial t}\rho_{ab} & = &
-i\Omega_D\rho_{ab}+i\lambda\rho_{aa}-\frac{1}{2}\Gamma\rho_{ab},\\
\frac{\partial}{\partial t}\rho_{ba} & = &
i\Omega_D\rho_{ba}+i\lambda\rho_{bb}-\frac{1}{2}\Gamma\rho_{ba}.
\end{eqnarray}
with the initial conditions $\rho_{ab}(0)=\rho_{ba}(0)=\rho_{aa}(0)=0$,
$\rho_{bb}(0)=1$.

Atom can act as an effective detector when the decay rate $\Gamma$ of the
excited state $|a\rangle$ is large. In such a situation we can obtain
approximate solution assuming that $\rho_{ba}$ and $\rho_{ab}$ are small and
$\rho_{bb}$ changes slowly. Then the approximate equation for the matrix
element $\rho_{ba}$ is
\begin{equation}
\frac{\partial}{\partial
t}\rho_{ba}=i\lambda\rho_{bb}(0)-\frac{1}{2}\Gamma\rho_{ba}
\end{equation}
with the solution
\begin{equation}
\rho_{ba}(t)=2i\frac{\lambda}{\Gamma}(1-e^{-\frac{1}{2}\Gamma t}).
\end{equation}
Substituting this solution into the equation for the matrix element $\rho_{bb}$
we get
\begin{equation}
\frac{\partial}{\partial
t}\rho_{bb}=-2\frac{\lambda^2}{\Gamma}(1-e^{-\frac{1}{2}\Gamma t}).
\end{equation}
Taking the term linear in $t$ we obtain the solution
\begin{equation}
\rho_{bb}(t)\approx1-2\frac{\lambda^2}{\Gamma}t\approx
e^{-2\frac{\lambda^2}{\Gamma}t}.
\label{eq:rhobb}
\end{equation}
Since $\rho_{aa}(t)=0$, using Eq.~(\ref{eq:rhobb}) we have
\[
F_{e,g}(t)=\rho_{bb}(t)\approx\exp\left(-\frac{t}{\tau_M}\right)
\]
where
\begin{equation}
\tau_M=\frac{\Gamma}{2\lambda^2}
\label{eq:tm}
\end{equation}
is the characteristic duration of the measurement. This estimate is justified
comparing with the exact solution of the equations. We get that the measurement
duration is shorter for bigger coupling strength $\lambda$.

\section{\label{sec:stochastic-methods}Stochastic methods}

The density matrix approach describes the evolution of a large ensemble of
independent systems. The observed signal allows us to generate an inferred
quantum evolution conditioned by a particular observed record
\cite{carmichael93}. This gives basis of the quantum jump models. In such
models the quantum trajectory is calculated by integrating the time-dependent
Schr\"odinger equation using a non-Hermitian effective Hamiltonian. Incoherent
processes such as spontaneous emission are incorporated as random quantum jumps
that cause a collapse of the wave function to a single state. Averaging over
many realizations of the trajectory reproduces the ensemble results.

The theory of quantum trajectories has been developed by many authors
\cite{carmichael93,dalibard92,dum92,gardiner92,gisin84,gisin92,schack95}.
Quantum trajectories were used to model continuously monitored open systems
\cite{carmichael93,dum92,gardiner92}, in the numerical calculations for the
study of dissipative processes \cite{dalibard92,schack95}, and in relation to
quantum measurement theory \cite{gisin84,gisin92}.

We assume that the Markovian approximation is valid. The dynamics of the total
system consisting of the measured system and the detector is described by a
master equation
\begin{equation}
\frac{\partial}{\partial
t}\hat{\rho}(t)=\mathcal{M}\hat{\rho},
\label{eq:master1}
\end{equation}
where $\mathcal{M}$ is the superoperator describing the time evolution. The
superoperator $\mathcal{M}$ can be separated into two parts
\begin{equation}
\mathcal{M}=\mathcal{L}+\mathcal{J}.
\label{eq:split}
\end{equation}
The part $\mathcal{J}$ is interpreted as describing quantum jumps,
$\mathcal{L}$ describes the jump-free evolution. After a short time interval
$\Delta t$ the density matrix is
\begin{equation}
\hat{\rho}(t+\Delta t)=\hat{\rho}(t)+\mathcal{L}\hat{\rho}(t)\Delta
t+\mathcal{J}\hat{\rho}(t)\Delta t.
\label{eq:m2}
\end{equation}
Since Eq.~(\ref{eq:master1}) should preserve the trace of the density matrix,
we have the equality
\begin{equation}
\Tr\{\mathcal{L}\hat{\rho}(t)\}+\Tr\{\mathcal{J}\hat{\rho}(t)\}=0.
\label{eq:tr}
\end{equation}
Using Eq.~(\ref{eq:tr}) equation (\ref{eq:m2}) can be rewritten in the form
\begin{equation}
\hat{\rho}(t+\Delta t)=\frac{\hat{\rho}(t)+\mathcal{L}\hat{\rho}(t)\Delta
t}{1+\Tr\{\mathcal{L}\hat{\rho}(t)\}\Delta
t}\left(1-\Tr\{\mathcal{J}\hat{\rho}(t)\}\Delta
t\right)+\frac{\mathcal{J}\hat{\rho}(t)}{\Tr\{\mathcal{J}\hat{\rho}(t)\}}\Tr\{
\mathcal{J}\hat{\rho}(t)\}\Delta t.
\label{eq:m3}
\end{equation}
This equation can be interpreted in the following way: during the time interval
$\Delta t$ two possibilities can occur. Either after time $\Delta t$ the
density matrix is equal to conditional density matrix
\begin{equation}
\hat{\rho}_{\mathrm{jump}}(t+\Delta
t)=\frac{\mathcal{J}\hat{\rho}(t)}{\Tr\{\mathcal{J}\hat{\rho}(t)\}}
\end{equation}
with the probability 
\begin{equation}
p_{\mathrm{jump}}(t)=\Tr\{\mathcal{J}\hat{\rho}(t)\}\Delta t
\end{equation}
or to the density matrix
\begin{equation}
\hat{\rho}_{\mathrm{no-jump}}(t+\Delta t)=\frac{\hat{\rho}(t)+\mathcal{L}\Delta
t\hat{\rho}(t)}{1+\Tr\{\mathcal{L}\hat{\rho}(t)\}\Delta t}
\end{equation}
with the probability $1-p_{\mathrm{jump}}(t)$. Thus the equation
(\ref{eq:master1}) can be replaced by the stochastic process.

Here we assume that the superoperators $\mathcal{L}$ and $\mathcal{J}$ have the
form
\begin{eqnarray}
\mathcal{L}\hat{\rho}& = &
\frac{1}{i\hbar}(\hat{H}_{\mathrm{eff}}\hat{\rho}-\hat{\rho}\hat{H}_{\mathrm{
eff}}^{\dag}),
\label{eq:L}
\\\mathcal{J}\hat{\rho}& = &
\hat{C}\hat{\rho}\hat{C}^{\dag}.
\label{eq:J}
\end{eqnarray}
The operators $\hat{H}_{\mathrm{eff}}$ and $\hat{C}$ are non-Hermitian in
general. If the superoperators $\mathcal{L}$ and $\mathcal{J}$ have the form
given in Eqs.~(\ref{eq:L}), (\ref{eq:J}) and the density matrix at the time $t$
factorizes as $\hat{\rho}(t)=|\Psi(t)\rangle\langle\Psi(t)|$ then after time
interval $\Delta t$ the density matrices $\hat{\rho}_{\mathrm{jump}}(t+\Delta
t)$ and $\hat{\rho}_{\mathrm{no-jump}}(t+\Delta t)$ factorize also. Therefore,
equation for density matrix (\ref{eq:m3}) can be replaced by the corresponding
equation for the state vectors. The state vector after time $\Delta t$ in which
a jump is recorded is given by
\begin{equation}
|\Psi_{\mathrm{jump}}(t+\Delta
t)\rangle=\frac{1}{\sqrt{\langle\Psi(t)|\hat{C}^{\dag}\hat{
C}|\Psi(t)\rangle}}\hat{C}|\Psi(t)\rangle.
\end{equation}
The probability of a jump occurring in the time interval $\Delta t$ is
\begin{equation}
p_{\mathrm{jump}}(t)=\langle\Psi(t)|\hat{C}^{\dag}\hat{C}|\Psi(t)\rangle\Delta
t.
\label{eq:p-jump}
\end{equation}
If no jump occurs, the state vector evolves according to the non-Hermitian
Hamiltonian $\hat{H}_{\mathrm{eff}}$,
\begin{equation}
|\Psi_{\mathrm{jump}}(t+\Delta
t)\rangle=\frac{1}{\sqrt{\langle\Psi(t)|\left(1+\frac{i}{\hbar}(\hat{H}_{
\mathrm{eff}}^{\dag}-\hat{H}_{\mathrm{eff}})\Delta
t\right)|\Psi(t)\rangle}}\left(1-\frac{i}{\hbar}\hat{H}_{\mathrm{eff}}\Delta
t\right)|\Psi(t)\rangle.
\end{equation}

Numerical simulation takes place over discrete time with time step $\Delta t$.
When the wavefunction $|\Psi(t_n)\rangle$ is given, the wavefunction
$|\Psi(t_{n+1})\rangle$ is determined by the following algorithm
\cite{carmichael93}:
\begin{enumerate}
\item evaluate the collapse probability $p_{\mathrm{jump}}(t_n)$ according to
Eq.~(\ref{eq:p-jump}) \item generate a random number $r_n$ distributed
uniformly on the interval $[0,1]$ \item compare $p_{\mathrm{jump}}(t_n)$ with
$r_n$ and calculate $|\Psi_c(t_{n+1})\rangle$ according to the rule
\begin{eqnarray*}
|\Psi(t_{n+1})\rangle &\sim &\hat{C}|\Psi_c(t_n)\rangle,\quad
p_{\mathrm{jump}}(t_n)<r_n,\\ |\Psi(t_{n+1})\rangle &\sim &
\exp\left(-\frac{i}{\hbar}\hat{H}_{\mathrm{eff}}\Delta
t\right)|\Psi(t_n)\rangle,\quad p_{\mathrm{jump}}(t_n)>r_n.
\end{eqnarray*}
\end{enumerate}
We can approximate the second case as
\[
|\Psi(t_{n+1})\rangle\sim\left(1-\frac{i}{\hbar}\hat{H}_{\mathrm{eff}}\Delta
t\right)|\Psi(t_n)\rangle.
\]

\section{\label{sec:stochastic-simulation}Stochastic simulation of the detector}

At first we consider the measurement of the unperturbed system and take the
perturbation $\hat{V}=0$. The measured system is an atom with the states
$|g\rangle$ and $|e\rangle$. The Hamiltonian of the measured atom is
\begin{equation}
\hat{H}_A=\hbar\omega_A|e\rangle\langle e|,
\end{equation}
where $\hbar\omega_A$ is the energy of the excited level.

The stochastic methods described in Sec.~\ref{sec:stochastic-methods} were used
to perform the numerical simulations of the measurement process. Using the
equation (\ref{eq:LD}) we take the operator $\hat{C}$ in Eq.~(\ref{eq:J})
describing jumps in the form
\begin{equation}
\hat{C}=\sqrt{\Gamma}\hat{\sigma}_{-}
\end{equation}
and the effective Hamiltonian in Eq.~(\ref{eq:L}) as
\begin{equation}
\hat{H}_{\mathrm{eff}}=\hat{H}_A+\hat{H}_D+\hat{H}_I-i\hbar\frac{\Gamma}{2}\hat{
\sigma}_{+}\hat{\sigma}_{-}.
\end{equation}
The wavefunction of the measured system and the detector is expressed in the
basis of the eigenfunctions of the Hamiltonians of the measured system and the
detector 
\begin{equation}
|\Psi\rangle=c_{ea}|e\rangle|a\rangle+c_{eb}|e\rangle|b\rangle+c_{
ga}|g\rangle|a\rangle+c_{gb}|g\rangle|b\rangle.
\label{eq:func-2level}
\end{equation}
The effective Hamiltonian produces the following equations for the coefficients
of the wave function $|\Psi\rangle$
\begin{eqnarray}
\dot{c}_{ea} & = &
-i\left(\omega_A+\frac{\Omega_D}{2}-i\frac{\Gamma}{2}\right)c_{ea},
\label{eq:38}
\\\dot{c}_{eb} & = &
-i\left(\omega_A-\frac{\Omega_D}{2}\right)c_{eb},\\\dot{c}_{ga} & = &
-i\lambda c_{gb}-i\frac{\Omega_D}{2}c_{ga}-\frac{\Gamma}{2}c_{ga},\\
\dot{c}_{gb} & = & -i\lambda c_{ga}+i\frac{\Omega_D}{2}c_{gb}.
\label{eq:41}
\end{eqnarray}
Equations (\ref{eq:38})--(\ref{eq:41}) are used in the numerical simulations to
describe the evolution between the jumps. After the jump in the detecting atom
the unnormalized wavefunction is
\begin{equation}
\hat{C}|\Psi\rangle=\sqrt{\Gamma}(c_{ea}|e\rangle|b\rangle+c_{
ga}|g\rangle|b\rangle).
\end{equation}
The jump occurs with the probability $p_{\mathrm{jump}}$ obtained from
Eq.~(\ref{eq:p-jump}),
\begin{equation}
p_{\mathrm{jump}}=\Gamma\Delta
t\frac{|c_{ea}|^2+|c_{ga}|^2}{|c_{ea}|^2+|c_{ga}|^2+|c_{eb}|^2+|c_{gb}|^2}.
\label{eq:Punp}
\end{equation}

\begin{figure}
\includegraphics{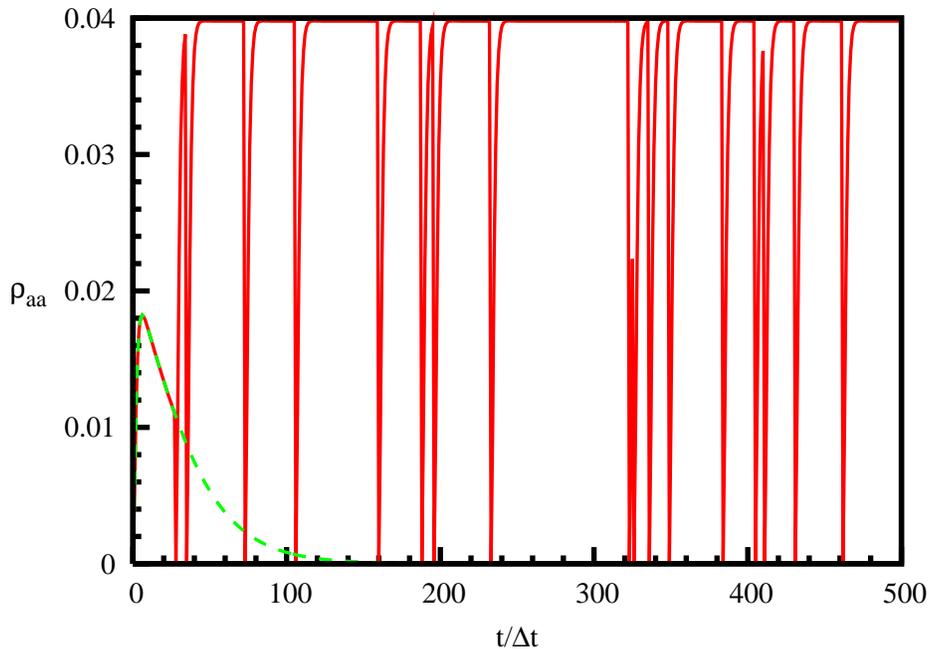}

\caption{Typical quantum trajectories of the detector. Figure shows the
probability $\rho_{aa}$ of being in the excited level of the detector. The
solid line corresponds to the case when the measured system collapses to the
ground state and the dashed line corresponds to the case when the measured
system collapses to the excited state. The parameters used for numerical
calculation are $\Delta t=0.1$, $\Gamma=10$, $\lambda=1$, and $\Omega_D=1$.}

\label{fig:trajekt-unpert}
\end{figure}

\begin{figure}
\includegraphics{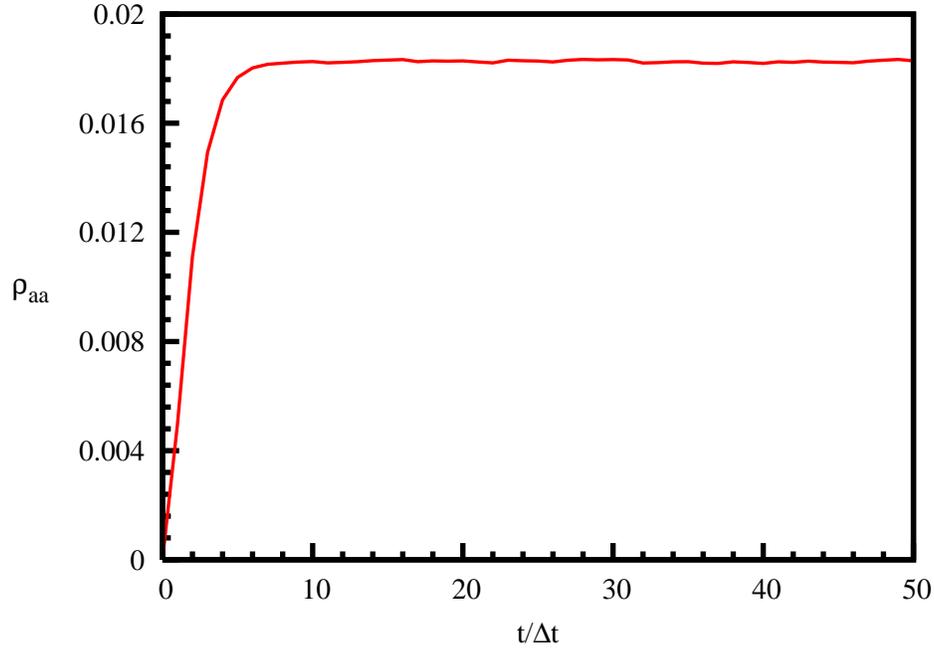}

\caption{Probability for the detector to be in the excited state, after
performing an ensemble average over 1000 trajectories. The parameters used are
the same as in Fig.~\ref{fig:trajekt-unpert}.}

\label{fig:aver-unpert}
\end{figure}

\begin{figure}
\includegraphics{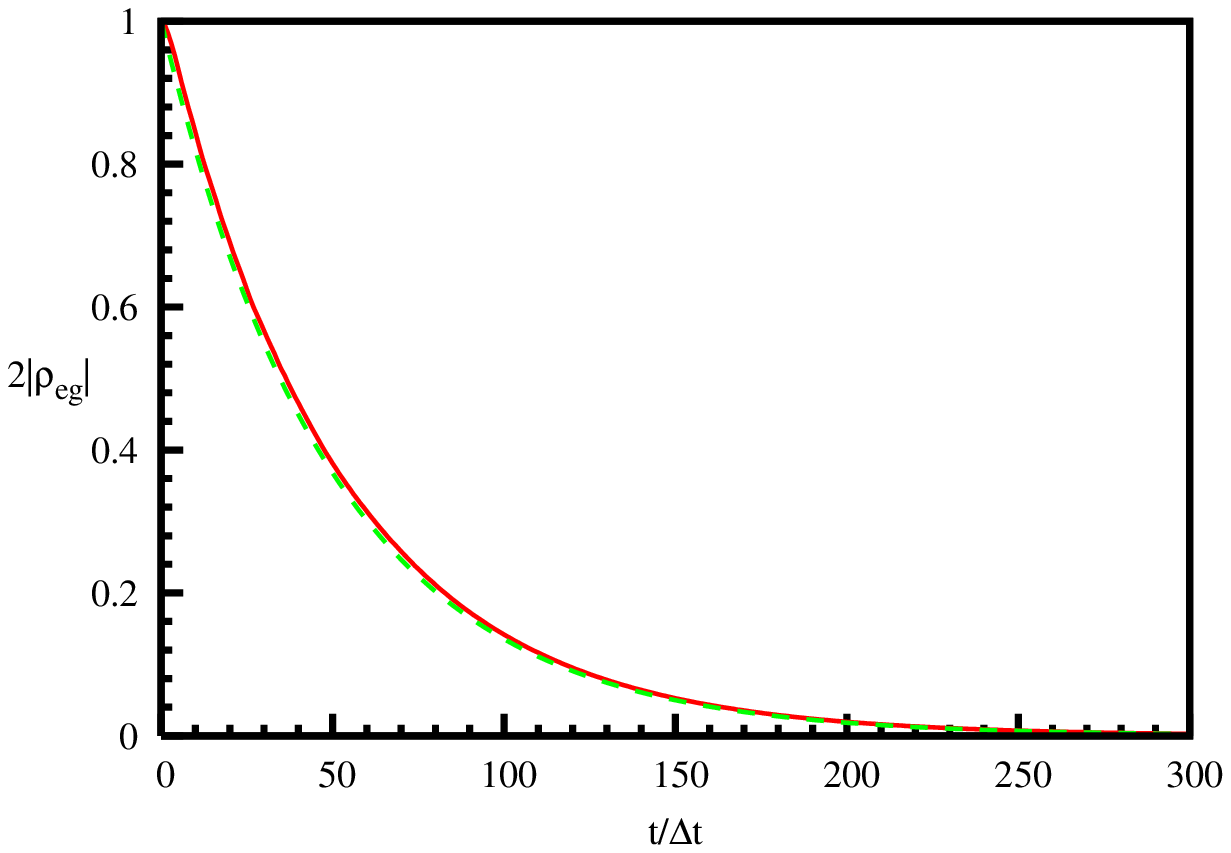}

\caption{Non-diagonal matrix elements of the density matrix of the measured
system. The solid line corresponds to the numerical calculations and the dashed
line corresponds to the exponential decay with the characteristic time given by
Eq.~(\ref{eq:tm}). The parameters used are the same as in
Fig.~\ref{fig:trajekt-unpert}.}

\label{fig:non-diag}
\end{figure}

For numerical simulation we take the measured system in an initial
superposition state $\frac{1}{\sqrt{2}}(|e\rangle+|g\rangle)$. The typical
quantum trajectories of the detector are shown in
Fig.~\ref{fig:trajekt-unpert}. There are two kinds of trajectories
corresponding to the collapse of the measured system to the excited or the
ground states. The trajectories corresponding to the collapse of the measured
system to the ground state show the repeated jumps. The mean interval between
jumps, obtained from the numerical simulation is of the same order of magnitude
as $\tau_M=5$ according to Eq.~(\ref{eq:tm}). After averaging over the
realizations the probability for the detector to be in the excited state is
shown in Fig.~\ref{fig:aver-unpert}. The figure shows that this probability
reaches the stationary value. The time dependency of the non-diagonal matrix
elements of the density matrix of the measured system is shown in
Fig.~\ref{fig:non-diag}. The figure shows a good agreement between the results
of numerical calculations and the exponential decay with the characteristic
time estimated from Eq.~(\ref{eq:tm}).

\section{\label{sec:measured-two-level}Frequently measured perturbed two level
system}

We consider an atom interacting with the classical external electromagnetic
field as the measured system. Interaction of the atom with the field is
described by the operator
\begin{equation}
\hat{V}=-\hbar\Omega_R(|e\rangle\langle g|+|g\rangle\langle e|)\cos\Omega t,
\end{equation}
where $\Omega$ is the frequency of the field and $\Omega_R$ is the Rabi
frequency. In the interaction representation and using the rotating-wave
approximation the perturbation $\hat{V}$ is
\begin{equation}
\tilde{V}(t)=-\hbar\frac{\Omega_R}{2}(e^{i\Delta\omega t}|e\rangle\langle
g|+e^{-i\Delta\omega t}|g\rangle\langle e|),
\label{eq:pert-rotating}
\end{equation}
where
\begin{equation}
\Delta\omega=\omega_A-\Omega
\end{equation}
is the detuning. Here $\hbar\omega_A=\hbar\omega_e-\hbar\omega_g$ is the energy
difference between the excited and the ground levels of the measured atom.

If the measurements are not performed, the atom exhibits Rabi oscillations with
the frequency $\Omega_R$. If the measured atom is initially in the state
$|g\rangle$, the time dependence of the coefficient $c_g$ of the wavefunction
$|\Psi\rangle=c_e|e\rangle+c_g|g\rangle$ is
\begin{equation}
c_g(t)=e^{-\frac{1}{2}it\Delta\omega}\left(\cos\left(\frac{1}{2}t\sqrt{
\Delta\omega^2+\Omega_R^2}\right)+i\frac{\Delta\omega}{\sqrt{\Delta\omega^2
+\Omega_R^2}}\sin\left(\frac{1}{2}t\sqrt{\Delta\omega^2
+\Omega_R^2}\right)\right).
\end{equation}
In particular, if the detuning $\Delta\omega$ is zero, we have
\begin{equation}
c_g(t)=c_g(0)\cos\left(\frac{\Omega_R}{2}t\right).
\end{equation}
When the measured atom interacts with the detector, we take the wavefunction of
the measured system and of the detector as in Eq.~(\ref{eq:func-2level}). In
the interaction representation the equations for the coefficients, when the
evolution is governed by the effective Hamiltonian $\hat{H}_{\mathrm{eff}}$,
defined by Eq.~(\ref{eq:L}), are
\begin{eqnarray}
\dot{c}_{ea} & = &
i\frac{\Omega_R}{2}e^{it\Delta\omega}c_{ga}-i\frac{\Omega_D}{2}c_{ea}-\frac{
\Gamma}{2}c_{ea}\\\dot{c}_{eb} & = &
i\frac{\Omega_R}{2}e^{it\Delta\omega}c_{gb}+i\frac{\Omega_D}{2}c_{eb}\\
\dot{c}_{ga} & = & i\frac{\Omega_R}{2}e^{-it\Delta\omega}c_{ea}-i\lambda
c_{gb}-i\frac{\Omega_D}{2}c_{ga}-\frac{\Gamma}{2}c_{ga}\\\dot{c}_{gb} & = &
i\frac{\Omega_R}{2}e^{-it\Delta\omega}c_{eb}-i\lambda
c_{ga}+i\frac{\Omega_D}{2}c_{gb}
\end{eqnarray}

The evolution of the measured atom significantly differs from the Rabi
oscillations. We are interested in the case when the duration of the
measurement $\tau_M$ is much shorter than the period of Rabi oscillations
$2\pi/\Omega_R$. In such a situation the non diagonal matrix elements of the
density matrix of the measured system remain small and the time evolution of
the diagonal matrix elements can be approximately described by the rate
equations
\begin{eqnarray}
\frac{d}{dt}\rho_{gg} & = &\Gamma_{e\rightarrow
g}\rho_{ee}(t)-\Gamma_{g\rightarrow e}\rho_{gg}(t),
\label{eq:aproxg}
\\
\frac{d}{dt}\rho_{ee} & = &\Gamma_{g\rightarrow
e}\rho_{gg}(t)-\Gamma_{e\rightarrow g}\rho_{ee}(t).
\label{eq:aproxe}
\end{eqnarray}
If the measured atom is initially in the state $|g\rangle$, the solution of
Eqs.~(\ref{eq:aproxg}) and (\ref{eq:aproxe}) is
\begin{equation}
\rho_{gg}(t)=\frac{\Gamma_{e\rightarrow g}+\Gamma_{g\rightarrow
e}e^{-(\Gamma_{e\rightarrow g}+\Gamma_{g\rightarrow e})t}}{\Gamma_{e\rightarrow
g}+\Gamma_{g\rightarrow e}}=\frac{1}{2}\left(1+e^{-2\Gamma_{g\rightarrow
e}t}\right).
\label{eq:solg}
\end{equation}
We can estimate the rates $\Gamma_{e\rightarrow g}$ and $\Gamma_{g\rightarrow
e}$ using equations from Ref.~\cite{ruseckas04}, i.e.,
\begin{eqnarray}
\Gamma_{e\rightarrow g} & = &
2\pi\int_{-\infty}^{\infty}G(\omega)P_{eg}(\omega)d\omega,
\label{eq:general-expr}
\\\Gamma_{g\rightarrow e} & = &
2\pi\int_{-\infty}^{\infty}G(\omega)P_{ge}(\omega)d\omega,
\end{eqnarray}
where
\begin{eqnarray}
P_{eg}(\omega)& = &
\frac{1}{\pi}\re\int_0^{\infty}F_{eg}(\tau)e^{i(\omega-\omega_A)\tau}d\tau\\ P_{
ge}(\omega)& = &
\frac{1}{\pi}\re\int_0^{\infty}F_{ge}(\tau)e^{i(\omega+\omega_A)\tau}d\tau
\end{eqnarray}
and
\begin{equation}
G(\omega)=\left(\frac{\Omega_R}{2}\right)^2[\delta(\omega-\omega_A
+\Delta\omega)+\delta(\omega+\omega_A-\Delta\omega)]\,.
\label{eq:Go2}
\end{equation}
Here the expression for $G(\omega)$ is derived using
Eq.~(\ref{eq:pert-rotating}). In contrast to Ref.~\cite{ruseckas04} in the
expression for $P(\omega)$ we extended the range of the integration to the
infinity since $F_{eg}(\tau)$ naturally limits the duration of the measurement.
Expressions, analogous to (\ref{eq:general-expr}), were obtained in Refs.
\cite{kofman00,kofman01}, as well.

Using Eqs.~(\ref{eq:general-expr})--(\ref{eq:Go2}) we can estimate the
transition rates as
\begin{equation}
\Gamma_{e\rightarrow g}\approx\Gamma_{g\rightarrow
e}\approx\frac{\Omega_R^2}{2}\re\int_0^{\infty}F_{eg}(\tau)e^{
-i\tau\Delta\omega}d\tau=\frac{\Omega_R^2}{2}\int_0^{\infty}e^{-\frac{\tau}{
\tau_M}-i\tau\Delta\omega}d\tau=\frac{\Omega_R^2}{2}\frac{\tau_M}{1
+(\tau_M\Delta\omega)^2}.
\label{eq:g-anti}
\end{equation}
Here we used the expression $\exp(-\tau/\tau_M)$ for $F_{eg}(\tau)$. When the
detuning $\Delta\omega$ is zero the transition rates are
\begin{equation}
\Gamma_{e\rightarrow g}\approx\Gamma_{g\rightarrow
e}\approx\frac{\Omega_R^2\tau_M}{2}.
\end{equation}
The transition rates are smaller for the shorter measurements. This is a
manifestation of the quantum Zeno effect.

\begin{figure}
\includegraphics{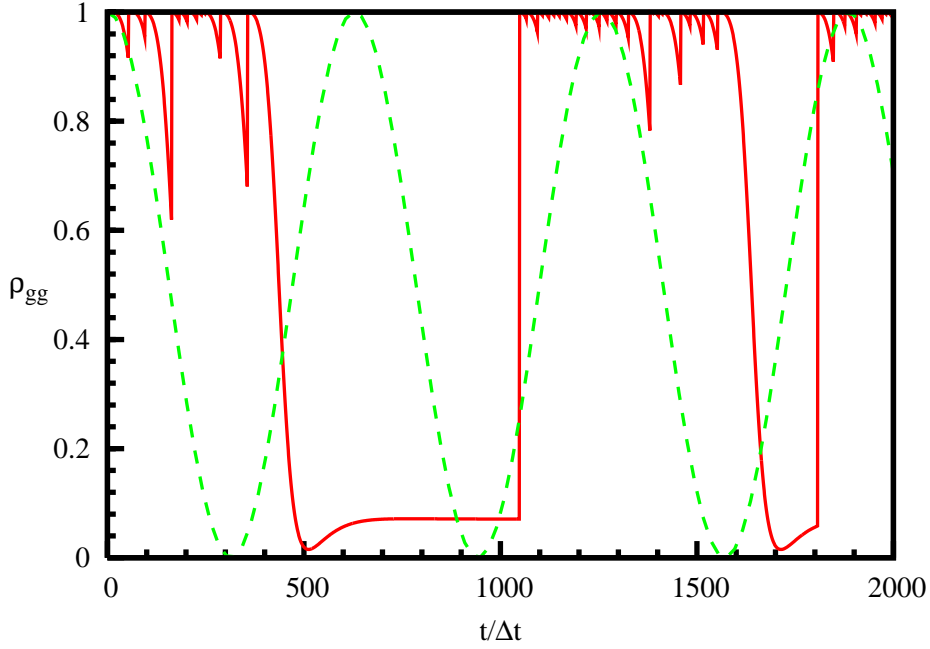}

\caption{Typical quantum trajectory of the measured two level system (solid
line). Figure shows the probability $\rho_{gg}$ for the measured atom to be in
the ground level. The dashed line shows the Rabi oscillations in the
measurement-free evolution. The detuning $\Delta\omega$ is zero. The parameters
used for the numerical calculation are $\Delta t=0.1$, $\Gamma=10$,
$\lambda=1$, $\Omega_D=1$, and $\Omega_R=0.1$.}

\label{fig:traekt-pert}
\end{figure}

\begin{figure}
\includegraphics{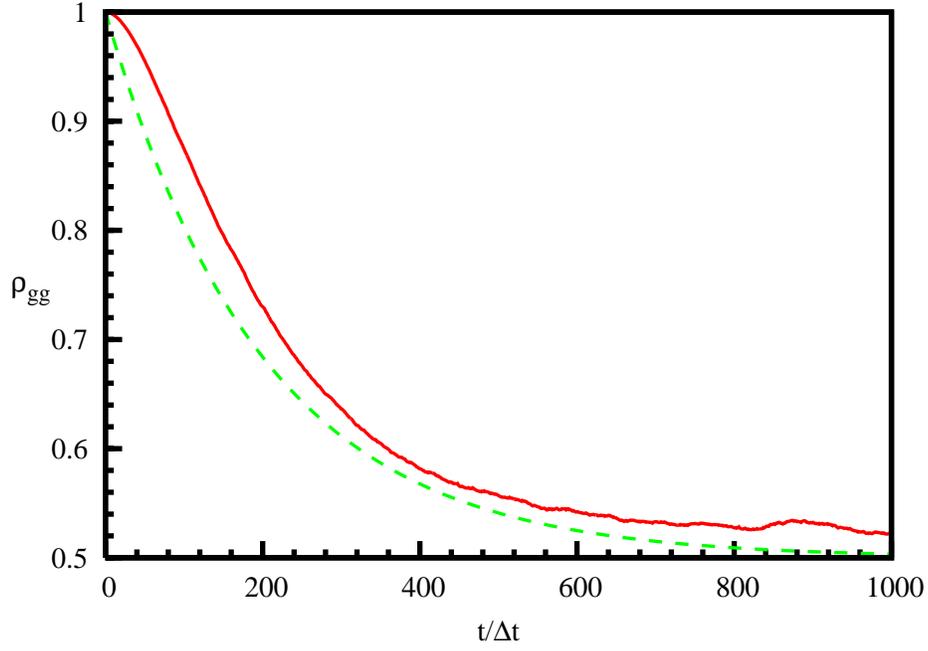}

\caption{Probability for the measured atom to be in the ground level, after
performing an ensemble average over 1000 trajectories. The solid line shows the
results of the numerical calculations, the dashed line shows the approximation
according to Eq.~(\ref{eq:solg}). The parameters used are the same as in
Fig.~\ref{fig:traekt-pert}.}

\label{fig:aver-pert}
\end{figure}

For numerical simulation we take the measured system in an initial state
$|g\rangle$. The typical quantum trajectory of the measured system is shown in
Fig.~\ref{fig:traekt-pert}. The behaviour of the measured system strongly
differs from the measurement-free evolution. The measurement-free system
oscillates with the Rabi frequency, while the measured system stays in one of
the levels and suddenly jumps to the other. The probability for the measured
atom to be in the ground state calculated after averaging over the realizations
is shown in Fig.~\ref{fig:aver-pert}. The figure shows that this probability
exhibits almost the exponential decay and after some time reaches the
stationary value close to $1/2$. The figure shows a good agreement between the
results of the numerical calculations and the estimate (\ref{eq:solg}).

\begin{figure}
\includegraphics{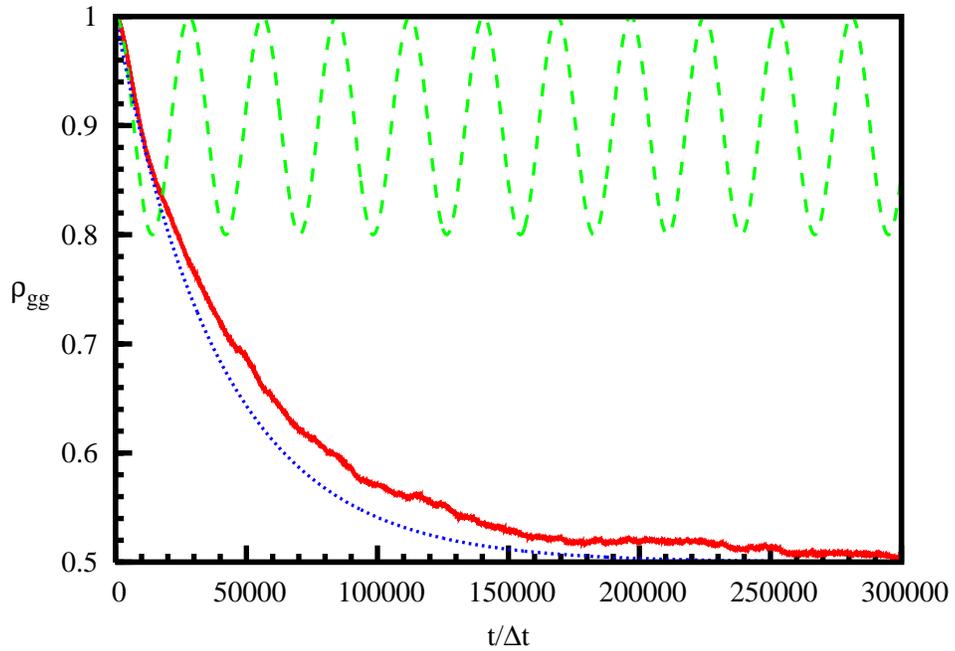}

\caption{Probability for the atom with the nonzero detuning to be in the ground
level. The solid line shows the results of the numerical simulation. The dotted
line shows the approximation according to Eq.~(\ref{eq:solg}), using the decay
rate from Eq.~(\ref{eq:g-anti}). The dashed line shows the evolution of a not
measured system. The parameters used for numerical calculation are $\Delta
t=0.001$, $\Gamma=10$, $\lambda=1$, $\Omega_D=1$, $\Omega_R=0.1$, and
$\Delta\omega=0.2$.}

\label{fig:2level-anti}
\end{figure}

When the detuning $\Delta\omega$ is not zero, the frequently measured two level
system can exhibit the anti-Zeno effect. This is pointed out in
Ref.~\cite{luis03}. For the case of nonzero detuning the probability that the
atom is in the ground state is shown in Fig.~\ref{fig:2level-anti}. The figure
shows that the probability for the atom to be in the initial (ground) state is
smaller, and, consequently, to be in the excited state is greater when the atom
is measured. This is the manifestation of the quantum anti-Zeno effect in the
two level system.

\section{\label{sec:decaying-system}Decaying system}

We model the decaying system as two level system $A$ interacting with the
reservoir $F$ consisting of many levels. The full Hamiltonian of the system is
\begin{equation}
\hat{H}=\hat{H}_A+\hat{H}_F+\hat{V},
\label{eq:61}
\end{equation}
where
\begin{equation}
\hat{H}_A=\hbar\omega_A|e\rangle\langle e|
\end{equation}
is the Hamiltonian of the two level system,
\begin{equation}
\hat{H}_F=\sum_k\hbar\omega_k|k\rangle\langle k|.
\end{equation}
is the Hamiltonian of the reservoir, and 
\begin{equation}
\hat{V}=\hbar\sum_k(g(k)|e\rangle\langle k|+g(k)^*|k\rangle\langle
e|)
\label{eq:64}
\end{equation}
describes the interaction of the system with the reservoir, with $g(k)$ being
the strength of the interaction with reservoir mode $k$. In the interaction
representation the perturbation $V$ has the form
\[
\tilde{V}(t)=e^{\frac{i}{\hbar}(\hat{H}_A+\hat{H}_F)t}\hat{V}e^{-\frac{i}{
\hbar}(\hat{H}_A+\hat{H}_F)t}=\hbar\sum_k(g(k)e^{i(\omega_A
-\omega_k)t}|e\rangle\langle k|+g(k)^*e^{-i(\omega_A
-\omega_k)t}|k\rangle\langle e|).
\]
The wavefunction of the system $A+F$ in the interaction representation is
expressed as
\begin{equation}
|\tilde{\Psi}\rangle=c_e(t)|e\rangle|0\rangle+\sum_kc_k|g\rangle|k\rangle.
\end{equation}
One can then obtain from the Schr\"odinger equation the following equations for
the coefficients
\begin{eqnarray}
\dot{c}_e & = &
-i\sum_kg(k)e^{i(\omega_A-\omega_k)t}c_k,
\label{eq:alpha}
\\\dot{c}_k
& = & -ig(k)^*e^{-i(\omega_A-\omega_k)t}c_e.
\label{eq:beta}
\end{eqnarray}
The initial condition is $|\Psi\rangle=|e\rangle|0\rangle$. Formally
integrating the equation (\ref{eq:beta}) we obtain the expression
\begin{equation}
c_k=-ig(k)^*\int_0^te^{-i(\omega_A-\omega_k)t'}c_e(t')dt'.
\label{eq:beta2}
\end{equation}
Inserting Eq.~(\ref{eq:beta2}) into Eq.~(\ref{eq:alpha}), we obtain the exact
integro-differential equation
\begin{equation}
\frac{d}{dt}c_e=-\int_0^tdt'\sum_k|g(k)|^2e^{i(\omega_A-\omega_k)(t-t')}c_e(t').
\label{eq:ce1}
\end{equation}
The sum over $k$ may be replaced by an integral
\[
\sum_k\rightarrow\int d\omega_k\rho(\omega_k)
\]
with $\rho(\omega_k)$ being the density of states in the reservoir. The
integration in Eq.~(\ref{eq:ce1}) can be carried out in the Weisskopf-Wigner
approximation. We get the equation
\begin{equation}
\frac{d}{dt}c_e=-\frac{\Gamma_{e\rightarrow g}^{(0)}}{2}c_e,
\end{equation}
where the decay rate $\Gamma_{e\rightarrow g}^{(0)}$ is given by the Fermi's
Golden Rule:
\begin{equation}
\Gamma_{e\rightarrow
g}^{(0)}=2\pi\rho(\omega_A)|g(\omega_A)|^2.
\label{eq:golden}
\end{equation}

In order to observe the quantum anti-Zeno effect one needs to have sufficiently
big derivative of the quantity $\rho(\omega)|g(\omega)|^2$. In such a case the
decay rate given by Fermi Golden Rule (\ref{eq:golden}) is no longer valid. The
corrected decay rate may be obtained solving Eq.~(\ref{eq:ce1}) by the Laplace
transform method \cite{lewenstein88}. The the Laplace transform of the solution
of Eq.~(\ref{eq:ce1}) is
\begin{equation}
\tilde{c}_e(z)=\frac{1}{\mathcal{H}(z)},
\end{equation}
where $\mathcal{H}(z)$ is the resolvent function
\begin{equation}
\mathcal{H}(z)=z+\int\frac{\rho(\omega)|g(\omega)|^2}{z+i(\omega
-\omega_A)}d\omega.
\end{equation}
In the numerical calculations we take the frequencies of the reservoir $\omega$
distributed in the region $[\omega_A-\Lambda,\omega_A+\Lambda]$ with the
constant spacing $\Delta\omega$. Therefore, the density of states is constant
$\rho(\omega)=1/\Delta\omega\equiv\rho_0$. The simplest choice of the
interaction strength $g(\omega)$ is to make it linearly dependent on $\omega$,
\begin{equation}
g(\omega)=g_0\left(1+\frac{a}{\Lambda}(\omega-\omega_A)\right),
\label{eq:g2}
\end{equation}
where $a$ is dimensionless parameter. Using Eq.~(\ref{eq:g2}) one obtains the
expression for the resolvent
\[
\mathcal{H}(z)=z+\pi\rho_0g_0^2\left[1-\frac{2}{\pi}\arctan\left(\frac{z}{
\Lambda}\right)+\left(a^2\frac{z}{\Lambda}-i2a\right)\left(\frac{2}{\pi}-\frac{
z}{\Lambda}+\frac{2}{\pi}\frac{z}{\Lambda}\arctan\left(\frac{z}{
\Lambda}\right)\right)\right].
\]
The real part of $z$ at which the resolvent $\mathcal{H}(z)$ is equal to zero
gives the decay rate. Expanding the resolvent into the series of powers of
$\Lambda^{-1}$ and keeping only the first-order terms we obtain the decay rate
\begin{equation}
\Gamma_{e\rightarrow g}^{(1)}\approx\Gamma_{e\rightarrow
g}^{(0)}\left(1-\frac{\Gamma_{e\rightarrow
g}^{(0)}}{\pi\Lambda}(5a^2-1)\right).
\label{eq:gamma-corr}
\end{equation}

\begin{figure}
\includegraphics{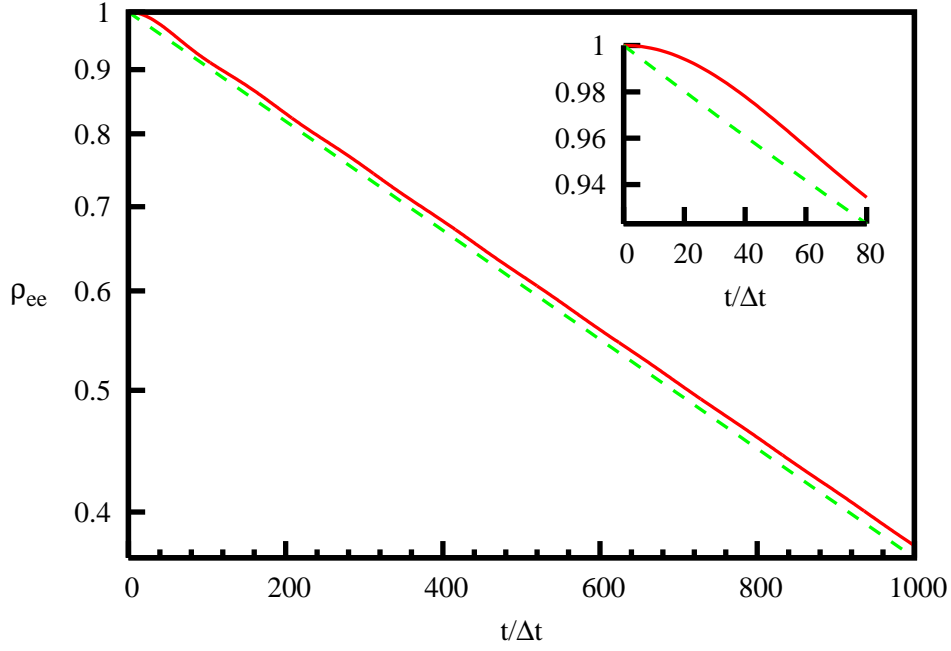}

\caption{Time dependence of the occupation of the exited level of the decaying
system. Solid line shows the results of the numerical calculation, dashed line
shows the exponential decay according to the Fermi's Golden Rule. The
parameters used for the numerical calculation are $\Delta t=0.1$,
$\Delta\omega=0.001$, $\Lambda=0.5$, and $g_0=0.001262$. For the parameters
used the decay rate is $\Gamma_{e\rightarrow g}^{(0)}=0.01$.}

\label{fig:decay-free}
\end{figure}

We solve Eqs.~(\ref{eq:alpha}) and (\ref{eq:beta}) numerically, replacing them
with discretized versions with the time step $\Delta t$. For calculations we
used $N=1000$ levels in the reservoir. The numerical results for constant
interaction strength are $g(k)=g_0$ presented in Fig.~\ref{fig:decay-free}. The
figure shows a good agreement between the numerical results and the exponential
law according to the Fermi's Golden Rule at intermediate times. At very short
time the occupation of the excited level exhibits quadratic behaviour, which,
for the repeated frequent measurements, may result in the quantum Zeno effect.

\begin{figure}
\includegraphics{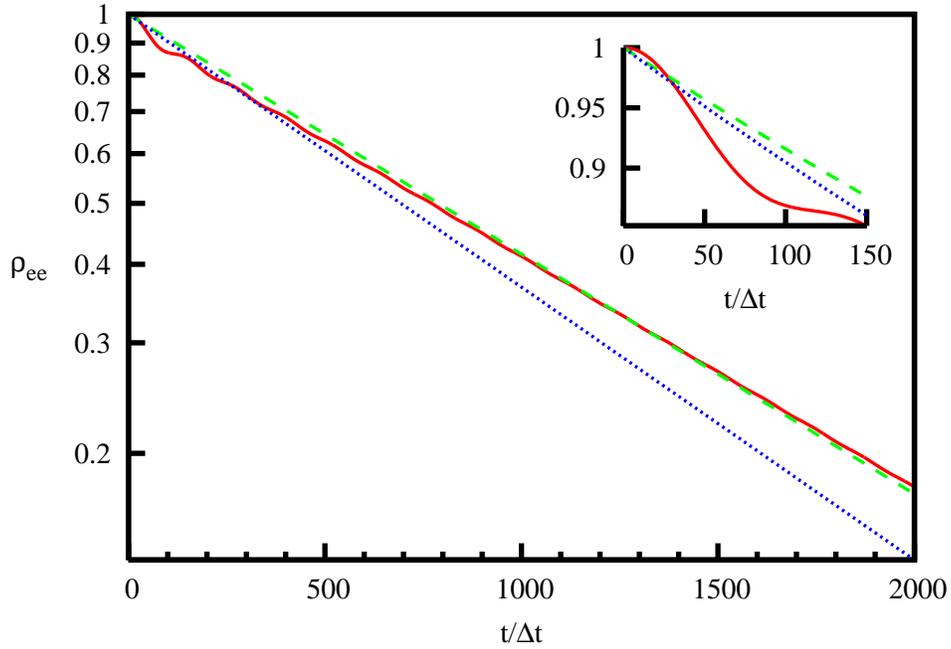}

\caption{Time dependence of the occupation of the exited level of the decaying
system when the interaction with the reservoir modes is described by
Eq.~(\ref{eq:g2}). Solid line shows results of numerical calculation, dashed
line shows exponential decay using the decay rate from
Eq.~(\ref{eq:gamma-corr}), dotted line shows exponential decay according to
Fermi's Golden Rule (\ref{eq:golden}). In the calculations we used $a=2$, other
parameters are the same as in Fig.~\ref{fig:decay-free}.}

\label{fig:decay-corr}
\end{figure}

Numerical results in the case when the interaction with the reservoir modes is
described by Eq.~(\ref{eq:g2}) with nonzero parameter $a$ are presented in
Fig.~\ref{fig:decay-corr}. The figure shows good agreement between the
numerical results and the exponential decay with the decay rate given by
Eq.~(\ref{eq:gamma-corr}) at intermediate times. For very short time we observe
the acceleration of the decay due to the interaction with the reservoir. This
acceleration for the repeated frequent measurements results in the quantum
anti-Zeno effect.

\section{\label{sec:measurement-decaying}Measurement of the decaying system}

In this section we consider the decaying system, described in
Sec.~\ref{sec:decaying-system} and interacting with the detector. The
wavefunction of the measured system and the detector, when the detector
interacts with the decaying system in the ground state only, we take in the form 
\begin{equation}
|\Psi_c\rangle=c_{ea}|e\rangle|0\rangle|a\rangle+c_{
eb}|e\rangle|0\rangle|b\rangle+\sum_k(c_{ka}|g\rangle|k\rangle|a\rangle+c_{
kb}|g\rangle|k\rangle|b\rangle).
\end{equation}
The equations for the coefficients, when the evolution is governed by the
effective Hamiltonian $\hat{H}_{\mathrm{eff}}$, are
\begin{eqnarray}
\dot{c}_{ea} & = &
-i\sum_kg(k)e^{i(\omega_A-\omega_k)t}c_{ka}-i\frac{\Omega_D}{2}c_{ea}-\frac{
\Gamma}{2}c_{ea},\\\dot{c}_{eb} & = &
-i\sum_kg(k)e^{i(\omega_A-\omega_k)t}c_{kb}+i\frac{\Omega_D}{2}c_{eb},\\\dot{
c}_{ka} & = & -ig(k)^*e^{-i(\omega_A-\omega_k)t}c_{ea}-i\lambda
c_{kb}-i\frac{\Omega_D}{2}c_{ka}-\frac{\Gamma}{2}c_{ka},\\\dot{c}_{kb} & = &
-ig(k)^*e^{-i(\omega_A-\omega_k)t}c_{eb}-i\lambda
c_{ka}+i\frac{\Omega_D}{2}c_{kb}.
\end{eqnarray}
After the jump in the detecting atom the unnormalized wavefunction becomes
\begin{equation}
\hat{C}|\Psi_c\rangle=\sqrt{\Gamma}(c_{ea}|e\rangle|0\rangle|b\rangle+\sum_kc_{
ka}|g\rangle|k\rangle|b\rangle).
\end{equation}

According to Ref.~\cite{ruseckas04}, the decay rate of the measured system is
given by expression (\ref{eq:general-expr}) with
\begin{equation}
G(\omega)=\rho(\omega)|g(\omega)|^2
\end{equation}
and
\begin{equation}
P(\omega)=\frac{1}{\pi}\re\int_0^{\infty}F_{eg}(\tau)e^{i(\omega
-\omega_A)\tau}d\tau.
\end{equation}
Using $F_{eg}(\tau)=\exp(-\tau/\tau_M)$ we obtain
\begin{equation}
P(\omega)=\frac{1}{\pi}\frac{\tau_M}{1+(\omega-\omega_A)^2\tau_M^2}.
\end{equation}
In order to obtain the quantum Zeno effect we take $G(\omega)$ as a constant
\begin{equation}
G(\omega)=\frac{\hbar^2g_0^2}{\Delta\omega},\quad\omega_A
-\Lambda\leq\omega\leq\omega_A+\Lambda.
\end{equation}
Here $\Delta\omega$ is the spacing between the modes of the reservoir. Using
Eq.~(\ref{eq:general-expr}) we get the decay rate of the measured decaying
system
\begin{equation}
\Gamma_{e\rightarrow g}=\Gamma_{e\rightarrow
g}^{(0)}\frac{2}{\pi}\arctan(\Lambda\tau_M).
\label{eq:gmma-meas}
\end{equation}
When $\Lambda\tau_M$ is big, we obtain the expression
\begin{equation}
\Gamma_{e\rightarrow g}=\Gamma_{e\rightarrow
g}^{(0)}\left(1-\frac{2}{\pi}\frac{1}{\Lambda\tau_M}+\cdots\right)
\label{eq:decay-zeno}
\end{equation}
by expanding Eq.~(\ref{eq:gmma-meas}) into series of the powers of
$(\Lambda\tau_M)^{-1}$. The second term in Eq.~(\ref{eq:decay-zeno}) shows that
the decay rate decreases with the decreasing duration of the measurement
$\tau_M$. This is the manifestation of the quantum Zeno effect.

\begin{figure}
\includegraphics{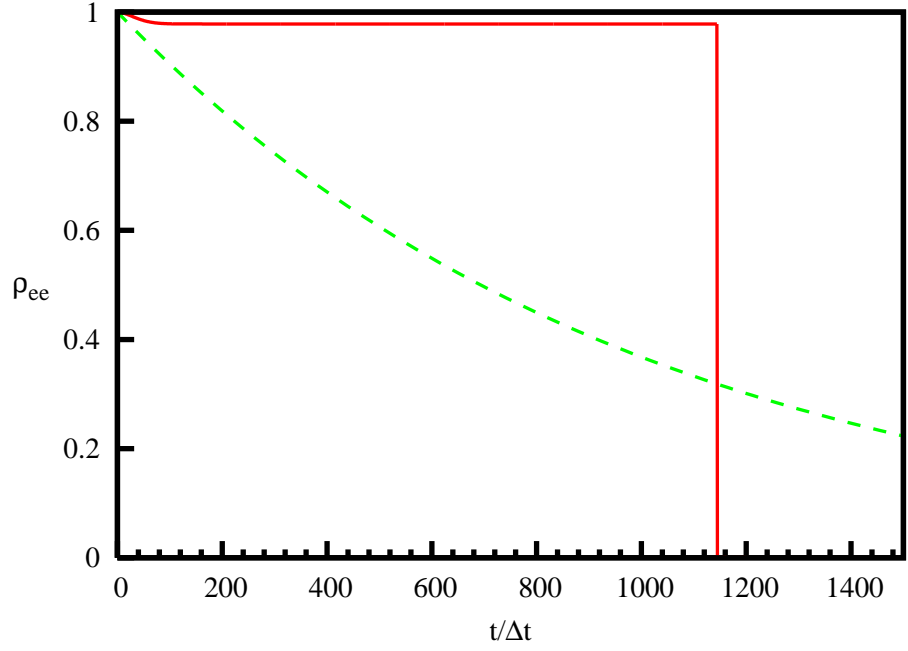}

\caption{Typical quantum trajectory (solid line) of the measured decaying
system when the detector interacts with the ground state of the measured
system. Figure shows the probability $\rho_{ee}$ that the measured atom is in
the excited level. The dashed line shows the exponential decay according to the
Fermi's Golden Rule in the measurement-free evolution. The parameters used for
the numerical calculation are $\Delta t=0.1$, $\Gamma=10$, $\lambda=1$,
$\Omega_D=1$, $\Delta\omega=0.001$, $\Lambda=0.5$, and $g_0=0.001262$. For the
parameters used the decay rate is $\Gamma_{e\rightarrow g}=0.01$.}

\label{fig:meas-dec-traj}
\end{figure}

\begin{figure}
\includegraphics{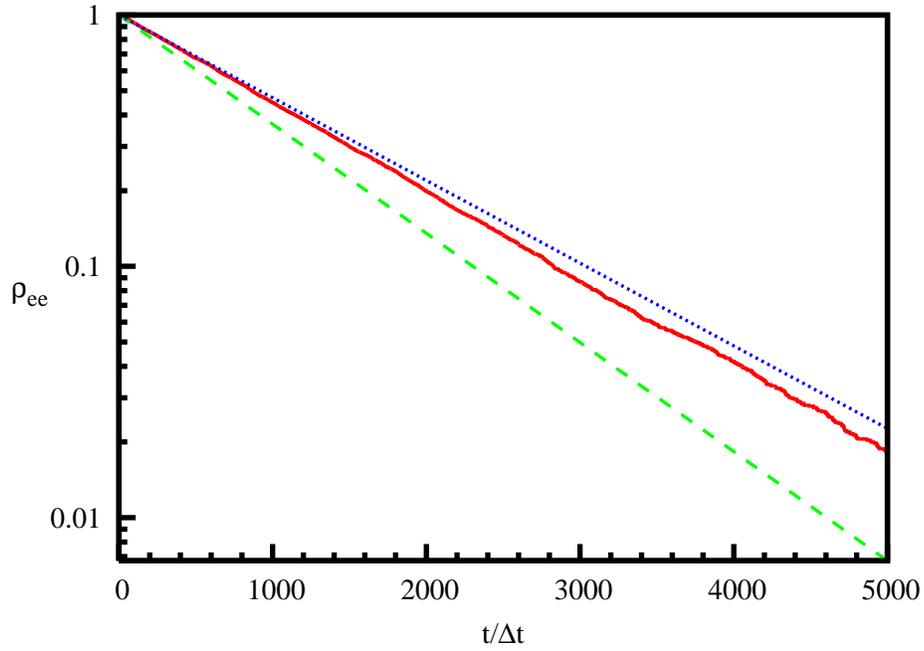}

\caption{Time dependence of the occupation of the exited level of the decaying
system. Solid line shows results of numerical calculation, dashed line shows
exponential decay according to Fermi's Golden Rule. The dotted line shows
approximation according to Eq.~(\ref{eq:gmma-meas}). The parameters used are
the same as in Fig.~\ref{fig:meas-dec-traj}.}

\label{fig:meas-decay}
\end{figure}

The results of the numerical simulation are presented in Figs.
\ref{fig:meas-dec-traj} and \ref{fig:meas-decay}. Typical quantum trajectory of
the measured decaying system is shown in Fig.~\ref{fig:meas-dec-traj}. This
trajectory is compatible with the intuitive quantum jump picture: the system
stays in the excited state for some time and then suddenly jumps to the ground
state. The probability that the measured system stays in the excited state is
presented in Fig.~\ref{fig:meas-decay}. Figure shows a good agreement between
the numerical simulation and the exponential law approximation with the
exponent given in Eq.~(\ref{eq:gmma-meas}). Also the quantum Zeno effect is
apparent.

\begin{figure}
\includegraphics{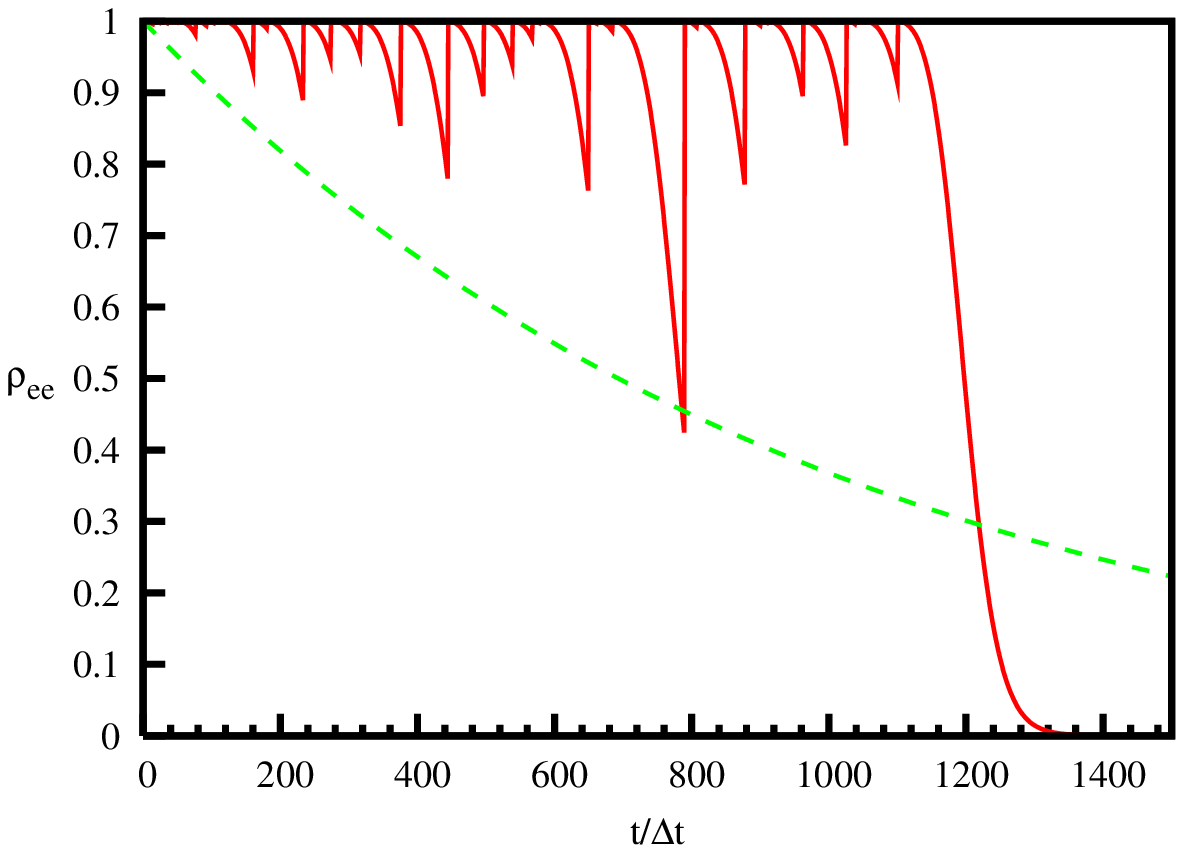}

\caption{Typical quantum trajectory of the measured decaying system (solid
line) when the detector interacts with the excited level. Figure shows the
probability $\rho_{ee}$ that the measured atom is in the excited level. The
dashed line shows exponential decay according to the Fermi's Golden Rule in the
measurement-free evolution. The parameters used are the same as in
Fig.~\ref{fig:meas-dec-traj}.}

\label{fig:meas-dec-traj-e}
\end{figure}

When the detector interacts with the excited state of the decaying system the
interaction term is
\begin{equation}
\hat{H}_I=\hbar\lambda|e\rangle\langle e|(\hat{\sigma}_{+}+\hat{\sigma}_{-})
\end{equation}
and the quantum trajectories are different. Typical quantum trajectory is shown
in Fig.~\ref{fig:meas-dec-traj-e}. This difference can be explained in the
following way: when the detector interacts with the ground state, the
interaction effectively begins only after some time, when the probability to
find the system in the ground state is sufficiently big. This explains the
absence of the collapses at short times in Fig.~\ref{fig:meas-dec-traj}. Then,
the measurement result after the collapse most likely will be that the system
is found in the ground state. When the detector interacts with the excited
state of the system, the interaction starts immediately and soon after that the
most probable result of the measurement is that the measured system is in the
excited state. It should be noted, that the averaged evolution shown in
Fig.~\ref{fig:meas-decay} does not depend on the state the detector is
interacting.

The model used for the decaying system when $g(\omega)=\mathrm{const}$ does not
allow to obtain the quantum anti-Zeno effect, since the conditions for the
quantum anti-Zeno effect, presented in Ref.~\cite{kofman00}, are not satisfied.
In order to obtain the quantum ant-Zeno effect we use the interaction with the
reservoir modes described by Eq.~(\ref{eq:g2}). In such a case we have
\begin{equation}
G(\omega)=\frac{\hbar^2g_0^2}{\Delta\omega}\left(1+\frac{a}{\Lambda}(\omega
-\omega_A)\right)^2,\quad\omega_A-\Lambda\leq\omega\leq\omega_A+\Lambda.
\label{eq:G2}
\end{equation}
Equation (\ref{eq:general-expr}) in this case does not give correct decay rate
of the measured system. In order to estimate the decay rate, we solve the
Liouville-von Neumann equation $i\hbar\dot{\rho}=[\hat{H},\rho]$ for the
density matrix of the system with the Hamiltonian (\ref{eq:61})--(\ref{eq:64}),
including additional terms describing decay of the non-diagonal elements with
rate $1/\tau_M$, i.e.,
\begin{eqnarray}
\dot{\rho}_{e0,e0} & = &
-i\sum_k(g(k)\rho_{gk,e0}-\rho_{e0,gk}g(k)^*),
\label{eq:ree}
\\
\dot{\rho}_{gk,gk'} & = &
-i\omega_{kk'}\rho_{gk,gk'}-i(g(k)^*\rho_{e0,gk'}-\rho_{gk,e0}g(k')),
\label{eq:rgg}
\\\dot{\rho}_{e0,gk} & = &
\left(-i(\omega_A-\omega_k)-\frac{1}{\tau_M}\right)\rho_{e0,gk}-i(\sum_{
k'}g(k')\rho_{gk',gk}-\rho_{e0,e0}g(k)),
\label{eq:reg}
\\\dot{\rho}_{gk,e0} & = &
\left(-i(\omega_k-\omega_A)-\frac{1}{\tau_M}\right)\rho_{gk,e0}-i(g(k)^*\rho_{
e0,e0}-\sum_{k'}\rho_{gk,gk'}g(k')^*).
\label{eq:rge}
\end{eqnarray}
We solve equations (\ref{eq:ree})--(\ref{eq:rge}) using the Laplace transform
method. Eliminating $\tilde{\rho}_{e0,gk}$ and $\tilde{\rho}_{gk,e0}$ from the
Laplace transform of Eqs. (\ref{eq:ree})--(\ref{eq:rge}) one gets the equations
for the Laplace transforms of the matrix elements of the density matrix,
\begin{eqnarray}
z\tilde{\rho}_{e0,e0}(z)-1 & = &
-\sum_k|g(k)|^2\left(\frac{1}{z+i(\omega_k-\omega_A)+\frac{1}{\tau_M}}+\frac{1}{
z+i(\omega_A-\omega_k)+\frac{1}{\tau_M}}\right)\tilde{\rho}_{
e0,e0}(z)\nonumber\\ & &
+\sum_{k,k'}\left(\frac{1}{z+i(\omega_k-\omega_A)+\frac{1}{\tau_M}}+\frac{1}{z
+i(\omega_A-\omega_{k'})+\frac{1}{\tau_M}}\right)\nonumber\\
& & \times g(k)g(k')^*\tilde{\rho}_{gk,gk'}(z),
\label{eq:rzee}
\\ (z+i\omega_{kk'})\tilde{\rho}_{gk,gk'}(z) & = &
-\sum_{k''}\left(\frac{g(k')g(k'')^*}{z+i(\omega_k-\omega_A)+\frac{1}{
\tau_M}}\tilde{\rho}_{gk,gk''}(z)+\frac{g(k'')g(k)^*}{z+i(\omega_A-\omega_{k'})
+\frac{1}{\tau_M}}\tilde{\rho}_{gk'',gk'}(z)\right)\nonumber\\ & &
+g(k')g(k)^*\left(\frac{1}{z+i(\omega_k-\omega_A)+\frac{1}{\tau_M}}+\frac{1}{z
+i(\omega_A-\omega_{k'})+\frac{1}{\tau_M}}\right)\tilde{\rho}_{e0,e0}(z).
\label{eq:rzgg}
\end{eqnarray}
On the r.h.s.\ of Eq.~(\ref{eq:rzgg}) we will neglect the small terms not
containing $\tilde{\rho}_{e0,e0}(z)$. Expressing $\tilde{\rho}_{gk,gk'}(z)$ via
$\tilde{\rho}_{e0,e0}(z)$ from Eq.~(\ref{eq:rzgg}), substituting into
Eq.~(\ref{eq:rzee}) and replacing the sum over $k$ by an integral we obtain
\begin{eqnarray}
\frac{1}{\tilde{\rho}_{e0,e0}(z)} & = & z+\int d\omega
G(\omega)\left(\frac{1}{z+i(\omega-\omega_A)+\frac{1}{\tau_M}}+\frac{1}{z
+i(\omega_A-\omega)+\frac{1}{\tau_M}}\right)\nonumber\\
& & -\int d\omega\int
d\omega'G(\omega)G(\omega')\frac{1}{z+i(\omega-\omega')}\nonumber\\
&&\times\left(\frac{1}{z+i(\omega-\omega_A)+\frac{1}{\tau_M}}
+\frac{1}{z+i(\omega_A-\omega')+\frac{1}{\tau_M}}\right)^2\,.
\label{eq:lapl-aze}
\end{eqnarray}
The value of $z$ at which the r.h.s of Eq.~(\ref{eq:lapl-aze}) is equal to zero
gives the decay rate. Using the expression (\ref{eq:G2}) for $G(\omega)$ and
keeping only the first-order terms of the expansion into series of the powers
of $\Lambda^{-1}$ we get the measurement-modified decay rate
\begin{equation}
\Gamma_{e\rightarrow g}=\Gamma_{e\rightarrow
g}^{(0)}\left(1-\frac{\Gamma_{e\rightarrow
g}^{(0)}}{\pi\Lambda}(5a^2-1)\right)+\Gamma_{e\rightarrow
g}^{(0)}\frac{2}{\pi}\frac{(a^2-1)}{\Lambda\tau_M}.
\label{eq:gamma-aze}
\end{equation}
Equation (\ref{eq:gamma-aze}) is valid only for sufficiently large duration of
the measurement $\tau_M$, since expansion into series requires that
$\Lambda\tau_M\gg1$. From Eq.~(\ref{eq:gamma-aze}) one can see that in order to
obtain the quantum anti-Zeno effect the parameter $a$ should be greater than
$1$. When the parameter $a$ is less than $1$ we get the Zeno effect, and when
$a=1$ the decay rate coincides with the decay rate of the free system.

\begin{figure}
\includegraphics{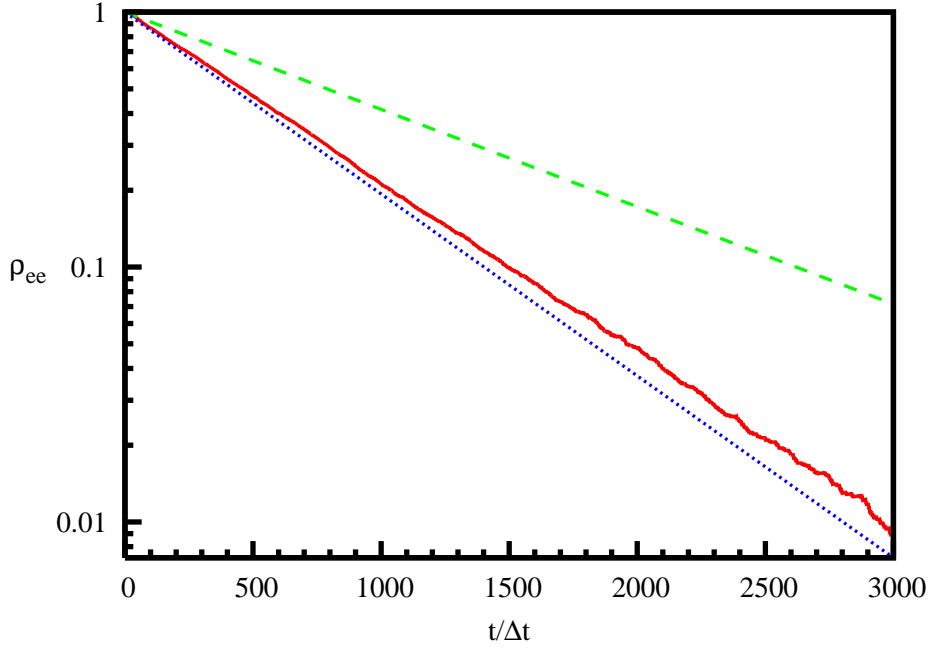}

\caption{Time dependence of the occupation of the exited level of the decaying
system when the interaction with the reservoir modes is described by
Eq.~(\ref{eq:g2}). Solid line shows results of the numerical calculation,
dashed line shows exponential decay of the measurement-free system with the
decay rate given by Eq.~(\ref{eq:gamma-corr}). The dotted line shows
approximation according to Eq.~(\ref{eq:gamma-aze}). In the calculations we
used $a=2$, while other parameters are the same as in
Fig.~\ref{fig:meas-dec-traj}.}

\label{fig:aze}
\end{figure}

The probability that the measured system stays in the excited state, obtained
from numerical simulation is presented in Fig.~\ref{fig:aze}. Figure clearly
demonstrates the quantum anti-Zeno effect, the decay rate of the measured
system is bigger than that of the measurement-free system.

\section{\label{sec:concl}Conclusions}

We model the evolution of the measured quantum system as a detector using a two
level system interacting with the environment. The influence of the environment
is taken into account using quantum trajectory method. The quantum trajectories
produced by stochastic simulations show the probabilistic behavior exhibiting
the collapse of the wave-packet in the measured system, although the quantum
jumps were performed only in the detector. Both quantum Zeno and anti-Zeno
effects were demonstrated for the measured two level system and for the
decaying system.

The results of the numerical calculations are compared with the analytical
expressions for the decay rate of the measured system. It is found that the
general expression (\ref{eq:general-expr}), obtained in Ref.~\cite{ruseckas04},
gives good agreement with the numerical data for the measured two level system
and for the decaying one showing the quantum Zeno effect. Nevertheless, when
the interaction of the measured system with the reservoir is strongly
mode-dependent, this expression does not give the correct decay rate. The decay
rate in this case was estimated including additional terms describing decay of
non-diagonal elements into the equation for the density matrix of the measured
system and a good agreement with the numerical calculations is found. A good
agreement of the numerical results with the analytical estimates of the decay
rates of the measured system shows that the particular model of the detector is
not important, since the decay rates mostly depend only on one parameter, i.e.,
the duration of the measurement (\ref{eq:tm}).


\end{document}